\documentclass[aps,pra,twocolumn,amsmath,amssymb,superscriptaddress,floatfix]{revtex4-1}

\usepackage{graphicx}
\usepackage{multirow}
\usepackage{color}
\usepackage{bm}
\usepackage{hyperref}

\def\t#1{\textrm{#1}}
\def\ket#1{|#1\rangle }

\def\braket#1{\langle #1 \rangle}
\def\n{\nonumber \\ }

\begin{document}

\title{
Nonreciprocal current from electron interactions in noncentrosymmetric crystals:
roles of time reversal symmetry and dissipation
}

\author{Takahiro Morimoto}
\affiliation{Department of Physics, University of California, Berkeley, CA 94720, USA}
\author{Naoto Nagaosa}
\affiliation{RIKEN Center for Emergent Matter Science (CEMS), Wako, Saitama, 351-0198, Japan}
\affiliation{Department of Applied Physics, The University of Tokyo, Tokyo, 113-8656, Japan}

\date{\today}

\begin{abstract}
\bf 
In noncentrosymmetric crystals with broken inversion symmetry 
$\mathcal{I}$, the 
$I-V$ ($I$: current, $V$: voltage) characteristic is generally expected 
to depend on the direction of $I$, which is known as
nonreciprocal response and, for example, found in p-n junction.
However, it is a highly nontrivial issue in translationally invariant systems 
since the time-reversal symmetry ($\mathcal{T}$) 
plays an essential role, where the two states at crystal momenta
$k$ and $-k$ are connected in the band structure. Therefore, it has been considered that
the external magnetic field ($B$) or the magnetic order which breaks the
$\mathcal{T}$-symmetry
is necessary to realize the nonreciprocal $I-V$ characteristics, i.e., magnetochiral 
anisotropy.   
Here we theoretically show that the electron correlation in $\mathcal{I}$-broken 
multi-band systems can induce nonreciprocal $I-V$ characteristics {\it without} 
$\mathcal{T}$-breaking. An analog of Onsager's relation shows that nonreciprocal current response without $\mathcal{T}$-breaking generally requires two effects: dissipation and interactions.
By using nonequilibrium Green's functions,
we derive general formula of the nonreciprocal response for two-band systems with onsite interaction. 
The formula is applied to Rice-Mele model, 
a representative 1D model with inversion breaking, and 
some candidate materials are discussed.
This finding offers a coherent understanding of the origin of 
nonreciprocal $I-V$ characteristics, and will pave a way to 
design it.
\end{abstract}

\maketitle

\begin{figure}[t]
\begin{center}
\includegraphics[width=0.98\linewidth]{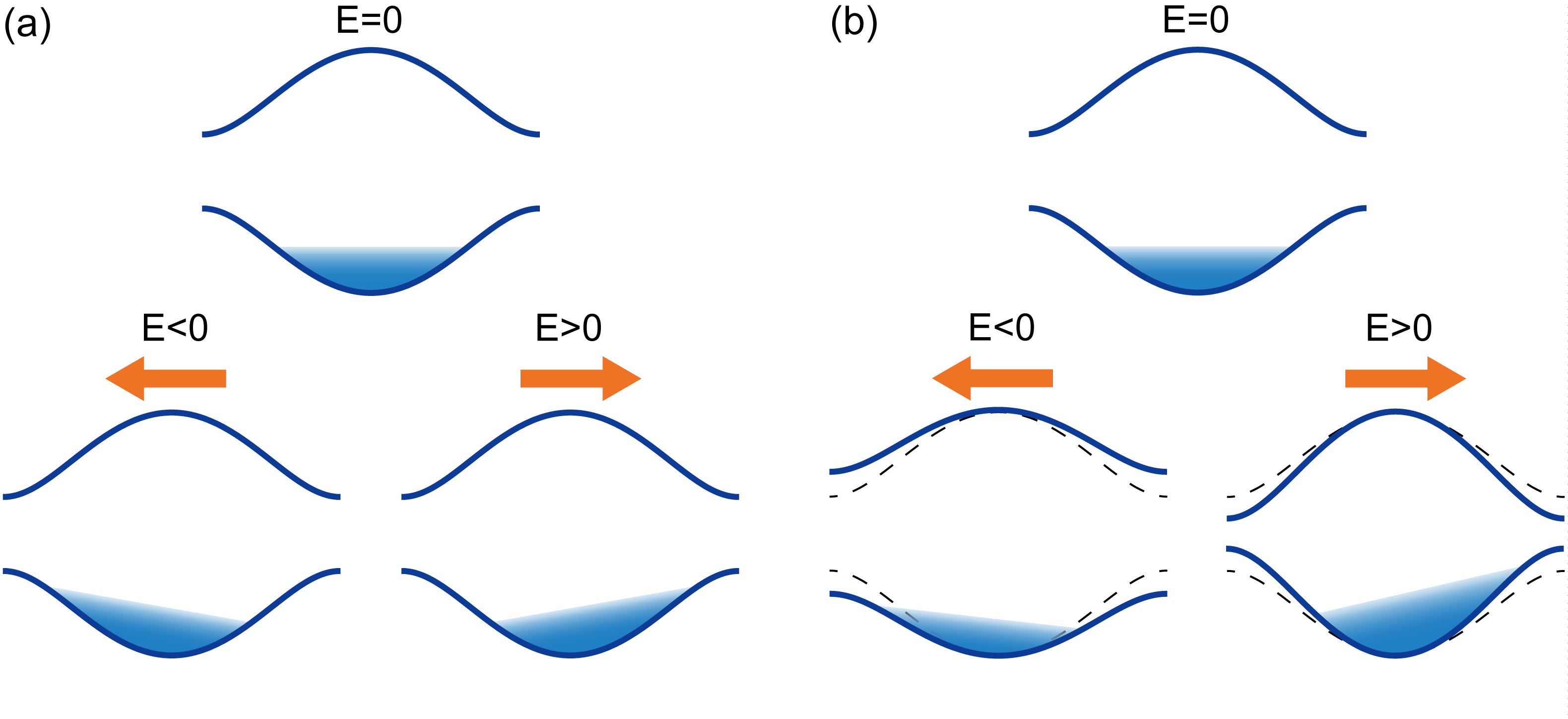}
\caption{\label{fig: schematics}
Schematic picture of the current responses in noncentrosymmetric crystals.
For simplicity, here we consider the spinless electrons.
(a) The conductivity of noninteracting electrons does not depend on the 
direction of the applied electric fields due to the time-reversal symmetry $\mathcal{T}$.
(b) Effective dispersion relation of interacting electrons are modified by the applied electric 
field $\bm E$ due to the electron correlation 
in a different way depending on its direction. This makes the conductivity 
 depend on the direction of $\bm E$, which is the nonreciprocal current response.
}
\end{center}
\end{figure}

Noncentrosymmetric crystals exhibit a variety of interesting physical phenomena. 
These include ferroelectricity \cite{Resta}, photovoltaic effect (shift current) 
\cite{Grinberg,Nie,Shi,deQuilettes,Kraut,Sipe,Young-Rappe,Cook17,Morimoto-shift16}, and second harmonic generation 
\cite{Boyd,Bloembergen,Wu16}. Among them, nonreciprocal dc current response in 
inversion broken systems has been attracting a keen attention in condensed matter 
physics. Nonreciprocity (or rectifying effect) is a current response where the $I-V$ characteristic 
differs when current flows toward left and when it flows toward right (i.e., $I(V) \neq -I(-V)$). The nonreciprocal 
current response is important both for fundamental physics of inversion broken materials and 
also for applications such as diode. Conventional example of nonreciprocity is a p-n junction, 
in which the direction of the current changes the thickness of depletion layer, and hence, 
the resistivity. 
Nonlinear current response has been intensely studied in a mesoscopic setup~\cite{ChristenButtiker96,Song99,SanchezButtiker04}.
In contrast to such artificial heterostructures, nonreciprocity 
in crystals is a more nontrivial issue. Current responses in crystals are governed by 
Bloch electrons with good momentum $k$ and their band structure.  
In the presence of time-reversal symmetry (TRS), the band structure satisfies the 
relationship $\epsilon_{k \sigma} = \epsilon_{-k {\bar \sigma}}$ ($\sigma$ represents the spin and
${\bar \sigma}$ the opposite spin to $\sigma$), 
which indicates that no nonreciprocity appears for noninteracting electrons in the 
Boltzmann charge transport picture as illustrated in Fig.~\ref{fig: schematics}(a). 
Specifically, the applied electric field causes a shift of the Bloch electrons in the 
momentum space. The symmetry in the band structure due to the TRS results in 
symmetric shifts with respect to the direction of the applied electric field $E$, 
and the conductivity does not depend on the direction of $E$. 
There are two ways to break TRS: (i) introducing time-reversal breaking term to a microscopic Hamiltonian and (ii) introducing irreversibility at the macroscopic level.
The former microscopic TR breaking is achieved
 with application of an external magnetic field $B$ or introducing magnetic order.
Nonreciprocal current response in the presence of magnetic field is known as magnetochiral anisotropy 
and has been actively studied 
\cite{RikkenNature,Rikken01,RikkenCNT,RikkenSi,Pop14,Morimoto-chiral16,Ideue,Wakatsuki}.

The other way to break TRS is incorporating irreversibility at the macroscopic level, i.e., (ii). Generalizing Onsager's relation to nonlinear current response, we find that nonreciprocal current may appear due to the effect of dissipation/relaxation even when microscopic Hamiltonian obeys the TRS. 
Furthermore, there exists a systematic description for the second order nonlinear current responses that is based on the gauge invariant formulation of nonequilibrium Green's functions under the static $E$ field~\cite{Onoda06,Sugimoto08}. This formulation allows us to show that some electron interaction effects are necessary for nonreciprocal current response in bulk crystals under the TRS, on top of the dissipation effects. Such electron interactions include Coulomb electrons between electrons and electron-phonon interactions. In particular, it turns out that elastic scattering from disorder potential is not able to support nonreciprocal current response.
These general symmetry considerations naturally lead us to study nonreciprocal current response with electron interactions in the Boltzmann transport picture that incorporates dissipation effects through relaxation of electron distribution functions.

Indeed, the situation changes in the presence of electron interactions, 
since electron interactions can modify the effective band structure when the applied electric field changes the electron distributions.
In the steady state with 
nonzero current in noncentrosymmetric crystals, the interaction effect modifies the energy band in an asymmetric way with respect to the direction of $E$, as illustrated in Fig.~\ref{fig: schematics}(b), and enables us to circumvent the original constraint of TRS 
since systems with $E$ and $-E$ are not related with TRS and $\epsilon_{k,\sigma}(E)
\neq \epsilon_{-k,\bar{\sigma}} (-E)$ in general.
In the Boltzmann transport picture, this asymmetric change of effective band structure leads to nonreciprocal current responses. 
By using the gauge invariant formulation of nonequilibrium Keldysh Green's functions, we derive a general formula for the nonreciprocal
current in the weak interaction limit.
It shows that nonreciprocal current in inversion broken materials is proportional to the strength of electron interaction 
and inversely proportional to typical band separation and bandwidth.
We find that 
the nonreciprocity in noncentrosymmetric crystals is a quantum mechanical effect that is 
described by the complex nature of Bloch wave functions and interband matrix element 
which is unique to inversion broken systems.
An estimate of the nonreciprocity shows that doped semiconductors and molecular conductors 
are good candidate materials for the nonreciprocity from electron correlation. 
We also discuss possible nonreciprocity in the molecular conductor TTF-CA \cite{Tokura88,Mitani87}. 

In this paper, we focus on the nonlinearity in I-V characteristic of dc transport. Meanwhile, there are other nonlinear current responses in $\mathcal{I}$-broken crystals which have their origins in the complex nature of Bloch wave functions and should be compared with the nonreciprocal response in the present case. 
One example is a shift current \cite{Kraut,Sipe,Young-Rappe,Cook17,Morimoto-shift16}, a dc current induced by photoexcitation of electrons beyond the band gap. The shift current is generated from the shift of wave packet centers for valence and conduction bands, and this shift is essentially described by Berry phases of valence and conduction electrons. 
The present nonreciprocal response and the shift current have similarity in that both rely on the multi-band nature of $\mathcal{I}$-broken systems. Yet, an important difference is that the nonreciprocal response arises from intraband metallic transport that is induced by static electric fields, while the shift current involves optical excitation of interband electron-hole pairs with photon energy larger than the band gap. 
Other examples are nonlinear Hall effect and low-frequency circular photogalvanic effect (CPGE)~ \cite{Moore-Orenstein10,Sodemann-Fu15,Morimoto-semiclassical16}. They are known as geometrical effects described by the Berry curvature dipole of Bloch electrons. They are similar to the present nonreciprocal response in that both are intraband effects.
However, the nonlinear Hall effect and the geometrical part of the CPGE are transverse (Hall) responses, in that they are described by off-diagonal components of the nonlinear conductivity tensor ($\sigma_{abb}$ and $\sigma_{aab}$, respectively, with $a \neq b$). In this sense, they are contrasted to the present nonreciprocal current which is a longitudinal current response described by diagonal components $\sigma_{aaa}$ and essentially involves the effect of dissipation.

\ \\
\par \ \
\noindent
\textbf{Results}
\par
Time reversal symmetry constrains nonreciprocal current responses in bulk crystals. 
Based on general symmetry considerations, we show that nonreciprocal current response in crystals generally require two ingredients: (i) dissipation, and (ii) interactions. First, we generalize Onsager's theorem to nonlinear current responses and show that the effect of dissipation is crucial for nonreciprocal current response. We then show by using gauge invariant formulation of Keldysh Green's function that nonreciprocal current generally requires some interactions (e.g., electron-electron interactions and electron-phonon interactions).
These two conditions suggest that the nonreciprocal current response can be captured by Boltzmann equation picture (that incorporates relaxation of electron distribution function) once we incorporate $E$-linear change of band structure induced by electron interactions.

The nonreciprocal current response is captured by an $E^2$ term in the current response. 
In the Boltzmann transport picture, 
the current $J$ induced by the applied electric field $E$ is given by
\begin{align}
J &= \frac{2e^2}{\hbar} \tau |v_F| E
\label{eq: boltzmann transport}
\end{align}
with the relaxation time $\tau$ and the Fermi velocity $v_F$,
 for a one-dimensional system as depicted in Fig.~\ref{fig: schematics}.
In noncentrosymmetric systems,
the effective band structure with correlation effect can change asymmetrically in an applied electric field, and the Fermi velocity is modified as $v_F(E)= v_{F,0} + c E + O(E^2) $.
Therefore, noncentrosymmetric systems can host nonreciprocal current response given by the $E^2$ term in $J= (2e^2/\hbar) \tau(v_{F,0}E+cE^2)$.
Since the $E$-linear change of the band structure is described by the self energy linear in $E$,
we study Green's function and self energy in the steady state realized with the
applied electric field. By using these results, we derive the general formula of nonreciprocal current, and then apply it to Rice-Mele model which is a prototypical model of ferroelectrics.

\textbf{Onsager's theorem and its generalization.}
In this section, we present a general consideration on the nonreciprocal current response 
in terms of the time reversal symmetry.
We generalize Onsager's relationship to nonlinear current responses,
and show that the effect of dissipation is crucial for nonreciprocal current response.

In the linear response, Onsager's relationship indicates that the conductivity $\sigma_{ij}$ 
is constrained as
\begin{align}
\sigma_{ij} &= \sigma_{ji},
\end{align}
when the microscopic Hamiltonian preserves time reversal symmetry  \cite{Onsager-reciprocal}.
This relationship is derived by considering the time reversal transformation in the Kubo formula 
for the linear conductivity as explained in Methods.
Now we study how Onsager's theorem can be extended to nonlinear 
current responses. We consider the second order current response,
\begin{align}
J_i(\omega_1 + \omega_2) &= \sigma_{ijj}(\omega_1, \omega_2) E_j(\omega_1)E_j(\omega_2).
\end{align}
For systems of noninteracting electrons, the nonlinear conductivity $\sigma_{ijj}(i\omega_{n_1},i\omega_{n_2})$ in the imaginary time formalism satisfies the relationship
\begin{align}
\sigma_{ijj}(i\omega_{n_1},i\omega_{n_2})
&=
-\sigma_{ijj}(-i\omega_{n_2},-i\omega_{n_1})
,
\label{eq: onsager nonlinear}
\end{align}
under time reversal symmetry.
(For the derivation, see Method section.)
Naively, this seems to suggest that the nonlinear conductivity $\sigma_{ijj}(\omega_1,\omega_2)$ 
vanishes in the dc limit ($\omega_1\to 0$ and $\omega_2\to 0$).
However, there is a subtlety in the analytic continuation to real frequencies as follows. We notice that nonlinear conductivity with Matsubara frequencies in the upper half plane is 
transformed to that with Matsubara frequencies in the lower half plane.
Since the real axis is a branch cut in the complex $\omega$ plane, the analytic continuation of 
$i\omega_n \to 0$ for the two quantities, $\sigma_{ijj}(i\omega_1,i\omega_2)$ and 
$\sigma_{ijj}(-i\omega_2,-i\omega_1)$, lead to different results in general.
This indicates that relationship similar to the Onsager's relation does not necessarily constrain the 
dc nonlinear conductivity to vanish.

Interestingly, the extended Onsager's relation in the above shows that nonreciprocal current response (nonzero $\sigma_{ijj}$) inevitably involves macroscopic irreversibility, i.e., the effect of dissipation, since the branch cut at $\t{Im}[\omega]=0$ is associated with macroscopic irreversibility. Specifically, such discontinuity for $\omega \to \pm 0 i$ appears in the self energy by incorporating dissipative processes such as impurity scattering. To see this, it is useful to consider the case of linear conductivity. Metallic conductivity $\sigma_{xx}(\omega)$ has a branch cut and the limit of $\omega \to +0i$ gives a dissipative current response which is proportional to the relaxation time $\tau$. In contrast, Hall conductivity $\sigma_{xy}(\omega)$ does not involve such branch cut and corresponds to nondissipative current response (independent of $\tau$).
Therefore, the nonreciprocal current response requires dissipation and should be proportional to the relaxation time $\tau$.

In passing, we note that Eq.~(\ref{eq: onsager nonlinear}) also indicates that the dissipation is essential for shift current which is a photocurrent caused by an optical resonance at a frequency $\omega$ above the band gap and described by $\sigma_{ijj}(\omega,-\omega)$ \cite{Nagaosa-Morimoto}. If there is no effect of dissipation, we can naively take analytic continuation of Eq.~(\ref{eq: onsager nonlinear}), which leads to $\sigma_{ijj}(\omega,-\omega)= - \sigma_{ijj}(\omega,-\omega)=0$. Thus nonzero shift current requires some irreversibility. This observation is coherent with the fact that shift current essentially relies on optical absorption which is an irreversible process.


\textbf{Absence of dc nonreciprocal current in noninteracting systems.}
In this section, we show that dc nonreciprocal current response does not appear when we do not incorporate effects of electron interactions that cause an effective change of the band structure under the applied electric field.
We first show that no nonreciprocal current response appears in periodic systems. We then generalize the proof to the systems with static disorder potentials and show that incorporating the effects of elastic scattering does not lead to nonreciprocal current response.

We study systems with an applied electric field by using Keldysh Green's 
function and its gradient expansion \cite{Rammer86,Jauho94,Kohler,Kamenev}. In particular, we use its gauge invariant formulation which enables us to treat the effect of $E$ directly \cite{Onoda06}. 
In the presence of a constant external electric field $E$, the Green's function and the self energy 
are expanded with respect to $E$ as \cite{Onoda06,Sugimoto08}
\begin{align}
G(\omega,k) &= G_0(\omega,k) + \frac{E}{2} G_E(\omega,k) + \frac{E^2}{8} G_{E^2}(\omega,k) +O(E^3), \\
\Sigma(\omega,k) &= \Sigma_0(\omega,k) + \frac{E}{2} \Sigma_E(\omega,k) + \frac{E^2}{8} \Sigma_{E^2}(\omega,k) +O(E^3),
\label{eq: G G0 GE}
\end{align}
where we set $\hbar=1,e=1$ for simplicity.
The unperturbed part of the Green's function $G_0$ is given by
\begin{align}
\begin{pmatrix}
G_0^R & G_0^K \\
0 & G_0^A
\end{pmatrix}^{-1}
&=
\omega-H -
\begin{pmatrix}
\Sigma_0^R & \Sigma_0^K \\
0 & \Sigma_0^A
\end{pmatrix},
\end{align}
with the unperturbed Hamiltonian $H$ (without $E$).
The linear order correction to the Green's function $G_E$ is given by
\begin{align}
G_E&=G_0 \Big[
\Sigma_E 
+ \frac{i}{2} 
\big( 
(\partial_{\omega} G_0^{-1}) G_0  (\partial_{k} G_0^{-1})
\n
& \hspace{5em} -
(\partial_{k} G_0^{-1}) G_0 (\partial_{\omega} G_0^{-1})
\big) \Big] G_0.
\label{eq:GE}
\end{align}
In order to describe the nonequilibrium steady state with applied electric fields, we suppose that the system is coupled to a heat bath. The coupling to the heat bath stabilizes the nonequilibrium electron distribution, and is incorporated through the self energy $\Sigma_0$ as 
$\Sigma_0^{R/A}(\omega)= \mp i \Gamma/2$ and $\Sigma_0^K(\omega)=i\Gamma f(\omega)$, where $\Gamma$ is the coupling strength and $f(\omega)$ is the Fermi distribution function (for details, see Methods) \cite{Rammer86,Jauho94}.

The second order current response is given by the expectation value,
\begin{align}
J_{E^2} = -i \int d\omega dk \t{tr}[v(k) G_{E^2}^<(\omega,k)].
\label{eq: JE2}
\end{align}
In order to show the absence of the second order current response in noninteracting systems, we set $E$-dependent self energy corrections to zero ($\Sigma_E=\Sigma_{E^2}=0$).
Here, we also assumed that the heat bath coupled to the system (and gives $\Sigma_0$) is large enough such that it is not modified with applying electric fields.
With vanishing $E$-dependent self energies, $G_{E^2}$ can be written as \cite{Onoda06}
\begin{align}
G_{E^2} &=
-\frac{i}{2} G_0 [(\partial_\omega G_0^{-1}) (\partial_k G_E) - (\partial_k G_0^{-1}) (\partial_\omega G_E) ] \n
& + \frac{1}{4}G_0 [(\partial_\omega^2 G_0^{-1}) (\partial_k^2 G_0) + (\partial_k^2 G_0^{-1}) (\partial_\omega^2 G_0) ].
\label{eq: ge2}
\end{align}

We can show that the expectation value $J_{E^2}$ vanishes in the presence of TRS as follows.
The TRS defined with $\mathcal T=K$ constrains Green's functions and velocity operator as
\begin{align}
G_0(\omega, k) &= G_0(\omega, -k)^T, \\
G_E(\omega, k) &= -G_E(\omega, -k)^T, \\
v(k) &= v(-k)^T, 
\end{align}
where $T$ denotes transposition with respect to the band index.
This transformation law leads to cancellation of the integrand of $J_{E^2}$ between $k$ and $-k$.
For example, the first term in $G_{E^2}$ in Eq.~(\ref{eq: ge2}) gives the contribution which transforms as
\begin{align}
&\t{tr}[v(k)G_0(\omega,k) (\partial_\omega G_0^{-1}(\omega,k)) (\partial_k G_E(\omega,k))] \n
&=
-\t{tr}[v^T(-k) G_0^T(\omega,-k) (\partial_\omega G_0^{-1,T}(\omega,-k)) (\partial_k G_E^T(\omega,-k))] \n
&= -\t{tr}[v(-k) G_0(\omega,-k) (\partial_\omega G_0^{-1}(\omega,-k)) (\partial_k G_E(\omega,-k))],
\end{align}
and cancels out between $k$ and $-k$. (In the last line, we used $\t{tr}A=\t{tr}A^T$.) We can show the cancellation for other terms in $J_E$ in a similar way.
This indicates that the nonlinear current $\propto E^2$ vanishes under the TRS in bulk crystals if we do not incorporate $E$-linear band modification described by $\Sigma_E$.

It is easy to generalize the above argument to systems with static disorder potential. We consider a system of the system size $L$ with the periodic boundary condition. We introduce a phase twist at the periodic boundary with the phase $\theta$. In this case, the velocity matrix element $v$ and the nonequilibrium Green's function $G_{E^2}$ become functions of the phase twist $\theta$ instead of the momentum $k$.
When the disorder is uniform and the system has translation symmetry on average, physical quantities are obtained by averaging over the phase twist $\theta$. We note that this procedure is very similar to the discussion of Chern number in quantum Hall systems with disorder potential \cite{Niu85}.
Thus, the nonlinear current response $J_{E^2}$ is given by
a similar expression to Eq.~(\ref{eq: JE2}) by replacing $k$ with $\theta$.
[The expression for $G_{E^2}(\omega,\theta)$ is also obtained by replacing $k$ with $\theta$ in Eq.~(\ref{eq: ge2}).]
Since similar symmetry constraints hold for $G$ and $v$ under the TRS [i.e., 
$G_0(\omega, \theta) = G_0(\omega, -\theta)^T,
G_E(\omega, \theta) = -G_E(\omega, -\theta)^T, 
v(\theta) = v(-\theta)^T
$
],
the integrand of $J_{E^2}$ satisfies 
\begin{align}
\t{tr}[v(\theta) G_{E^2}^<(\omega,\theta)]=- \t{tr}[v(-\theta) G_{E^2}^<(\omega,-\theta)],
\end{align}
and cancels between $\theta$ and $-\theta$.
This proves that elastic scattering from static disorder potential does not induce nonreciprocal current response.

These considerations indicate that $E$-linear change of band structure ($\Sigma_E$) is essential for nonreciprocal current response in bulk crystals.
The $E$-linear change of band structure requires some kind of electron interactions, such as Coulomb interactions and electron-phonon interactions.
Since the current response proportional to $E^2$ arises from the $E$-linear change of band structure in the Boltzmann transport picture,
it suffices to consider $\Sigma_E$ and neglect $\Sigma_{E^2}$.
Although we can study this nonreciprocal current response by directly looking at $G_{E^2}$ with incorporating $\Sigma_E$, it is equivalent and more concise to compute $\Sigma_E$ and then use the relationship Eq.~(\ref{eq: boltzmann transport}) with the Fermi velocity modified by $E$.

\begin{figure}[tb]
\begin{center}
\includegraphics[width=0.8\linewidth]{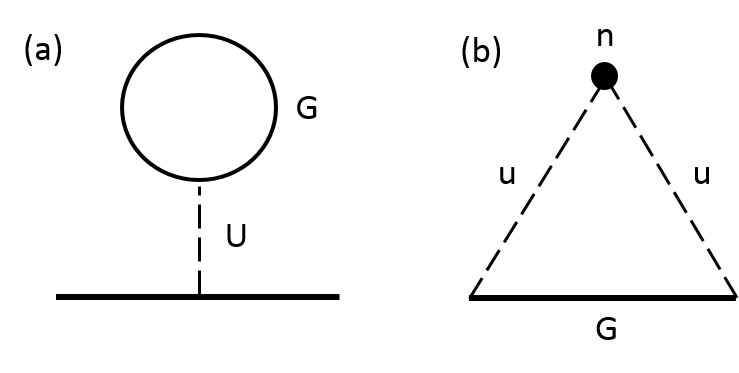}
\caption{\label{fig:diagrams}
Diagrams that we consider for (a) electron-electron interaction, (b) impurity scattering.
The electron-electron interaction is incorporated by the Hartree term.
We use the Born approximation (the second order perturbation) for the impurity scattering. 
}
\end{center}
\end{figure}

So far, we discussed general conditions to achieve nonreciprocal current response in bulk crystals. In order to proceed to explicit calculations of nonreciprocal current, we need to specify the form of the self energy, i.e., how the self energy $\Sigma$ is expressed in terms of the Green's function $G$.
We consider electron-electron interaction shown in the Feynman diagram Fig.~\ref{fig:diagrams}(a) and show that it gives rise to nonreciprocal current through $E$-linear band structure change. Incidentally, we also show explicitly that elastic scatterings from isotropic impurity potential [Fig.~\ref{fig:diagrams}(b)] does not lead to nonreciprocal current, which is consistent with the above general symmetry consideration.

\textbf{Nonequilibrium steady state under the applied electric field.}
Now we move on to demonstration of nonreciprocal current responses with electron interactions by performing explicit calculations. We consider the cases of weak interactions and perform Hartree-Fock approximation in the gauge invariant formulation of Keldysh Green's functions. 
In order to describe $E$-linear change of the effective band structure, we first study the nonequilibrium steady state under the electric field by looking at $G_E^<$. Once $G_E^<$ is obtained, we can compute the $E$-linear change of band structure by studying $\Sigma_E^R$ that corresponds to the diagram in Fig.~\ref{fig:diagrams}.

The $E$-linear change of electron occupation has intraband and interband contributions, since the Green's function for a $\mathcal{I}$-broken system generally has a matrix structure with respect to band index.
The intraband contribution is written as
\begin{align}
G_{E,11}^< &= \frac{2\pi i}{\Gamma} \delta(\omega-\epsilon_F) \sum_{k_{F,i}}\t{sgn}(v_{k_{F,i}}) 
\delta(k-k_{F,i}),
\label{eq:GElesser}
\end{align}
for the band 1 that we assume crosses with the Fermi energy (for details, see Methods).
Here, $k_{F,i}$ are the Fermi momenta for the band 1.
This change of the lesser Green's function linear in $E$ describes the effect of the applied electric field 
where the electron occupation is shifted in the momentum space as $k \to k + \tau E$ near the Fermi 
surface (with $\tau = 2\pi/\Gamma$).
This coincides with the picture of the semiclassical Boltzmann equation as illustrated in 
Fig.~\ref{fig: schematics}(a).

Next, the interband contribution for $G_E^<$ is given by
\begin{align}
G_{E,12}^< &=
- \sum_{k_{F,i}}
\frac{\pi v_{12,k}}{|v_{11,k}| E_{g,k}} \delta(\omega-\epsilon_F) \delta(k-k_{F,i})
,
\label{eq:Glesser-inter}
\end{align}
and $G_{E,21}^<=-(G_{E,12}^<)^*$,
for the bands 1 and 2 (for details of the derivation, see Methods.).
Here, we assume that the band 1 is the partially filled valence band and the band 2 is the unoccupied conduction band as illustrated in Fig.~\ref{fig: schematics}, and $E_{g,k}$ denotes the band gap at the momentum $k$.
This term arises from a quantum mechanical effect that the electric field also modifies the wave function in addition to the shift of the momentum at the Fermi energy. Thus the electron distribution in the steady state effectively has an interband component near the Fermi energy.
We note that this interband component of $G^<_E$ cannot be captured by semiclassical treatment 
with Boltzmann equation, and is a quantum effect captured by the current approach that uses
the gauge invariant formulation of Keldysh Green's functions.
This interband component gives the origin of the nonreciprocity when the electron interaction is incorporated.
In contrast, when we consider effects of scattering by short-range impurities within the Born approximation [described by the diagram in Fig.~\ref{fig:diagrams}(a)],
we do not find the $E$-linear change of the effective band structure,  as detailed in Methods.

\textbf{Formula of nonreciprocal current in two band systems.}
Now we show that nonreciprocal current appears from $E$-linear band structure change once we introduce electron-electron interactions, and derive a general formula for nonreciprocal current in two-band systems.
The effect of electron interactions is minimally incorporated by the self energy arising from the Hartree contribution to $\Sigma_E^R$ 
as shown in Fig.~\ref{fig:diagrams}(a).

For simplicity, we consider a two-band model, where
the unit cell contains two sites,
and the wave functions of valence and conduction bands (labeled by $1$ and $2$, respectively) are represented by 
\begin{align}
\Psi_{1,k} &= 
\begin{pmatrix}
u_k \\ v_k
\end{pmatrix},
&
\Psi_{2,k} &=
\begin{pmatrix}
-v^*_k \\ u^*_k 
\end{pmatrix}
.
\label{eq: Psi k}
\end{align}
We then consider two copies of the original system, each labeled by $\uparrow$ and $\downarrow$, and introduce the onsite interaction given by
\begin{align}
H_\t{int} &= U \sum_i n_{\uparrow,i} n_{\downarrow,i},
\label{eq: Hint}
\end{align}
with the site index $i$.
We treat the effects of the onsite interaction in terms of Hartree-Fock approximation, and study the effective band structure.
Since the two copies ($\uparrow$ and $\downarrow$) are decoupled in the noninteracting Hamiltonian, 
only the Hartree term appears in the present case.
(We suppose that the Hartree correction in the equilibrium is already included in the original Hamiltonian.)
In the following, we focus on the electronic structure of the $\uparrow$ component, and suppress the label for the two copies for simplicity.
By using the momentum space representation of $H_\t{int}$ 
(for details, see
Methods),
the self energy from the Hartree contribution is given by
\begin{align}
\Sigma_{E,11}^R(k)
&=
i a U \int \frac{d\omega}{2\pi} \frac{dk'}{2\pi} (|u_k|^2-|v_k|^2) \n
&\hspace{2.5em}
\times [ u_{k'}v_{k'} G_{E,12}^<(k') + u_{k'}^* v_{k'}^* G_{E,21}^<(k') ],
\label{eq: Sigma E 11 R}
\end{align}
with the lattice constant $a$.
Now we assume that there are two Fermi momenta at $\pm k_F$ with the same 
Fermi velocity $v_F$.
By using the Green's function in the steady state [Eq.~(\ref{eq:Glesser-inter})], this is expressed as
\begin{align}
\Sigma_{E,11}^R(k)
&=
\frac{a U (|u_k|^2-|v_k|^2)}{\pi |v_{11,k_F}| E_{g,k_F}} 
\t{Im}\left[u_{k_F}v_{k_F} v_{12,k_F}\right].
\end{align}
This self energy is an even function with respect to $k$ from TRS (such as $\mathcal{T}=\mathcal{K}$),
which is important in obtaining nonreciprocal current response as we will see next.

We now study the nonreciprocal current response by using the self energy $\Sigma_E^R$.
The current induced by an electric field (linearly in $E$) is given by
\begin{align}
J&=(v_{11,k_F} - v_{11,-k_F}) \tau E,
\label{eq: linear J}
\end{align}
from the Boltzmann transport approach.
An application of the electric field modifies the band structure as 
$\epsilon_1 \to \epsilon_1 +\frac{E}{2} \Sigma_{E,11}^R(k)$, and hence,
the Fermi velocity as $v_{11,k_F} \to v_{11,k_F} +\frac{E}{2} \partial_k \Sigma_{E,11}^R(k)$.
Since the obtained self energy $\Sigma_{E,11}^R(k)$ is an even function of $k$,
the velocity corrections at $\pm k_F$ do not cancel out in evaluating the correction to the 
current response in Eq.~(\ref{eq: linear J}).
Thus, we obtain the nonlinear current response $\delta J$ (the part of current response proportional to $E^2$) as
\begin{align}
\delta J &= \left(\left. \partial_k \Sigma_{E,11}^R(k) \right|_{k=k_F} - \left. \partial_k 
\Sigma_{E,11}^R(k) \right|_{k=-k_F} \right) \tau E^2
\n
&= \frac{2 a U \tau \left. \partial_k(|u_k|^2-|v_k|^2) \right|_{k=k_F} }{\pi |v_{11,k_F}| E_{g,k_F}} 
\t{Im}\left[u_{k_F}v_{k_F} v_{12,k_F}\right]
 E^2,
\label{eq: nonlinear delta J}
\end{align}
which is the general formula for two-band systems in one-dimension.
This can be generalized to systems in higher dimensions if we replace the summation over the Fermi points with an integral over the Fermi surface.
The above formula indicates that the nonreciprocity ratio $\gamma$ of the nonlinear current to the original current is roughly estimated as
\begin{align}
\gamma \equiv \frac{\delta J}{J} \simeq 
\frac{U}{E_{g,k_F}} \frac{eEa}{W},
\label{eq: nonreciprocity ratio}
\end{align}
where $W$ is the band width. Here we used  
$u_k, v_k \sim 1$ and $v_{11,k_F} \sim v_{12,k_F}$
for rough order estimates.

The obtained formula indicates that breaking of inversion symmetry is essential for the nonreciprocity.
When the system is inversion symmetric, the wave function is expressed with real numbers due to 
the combination of inversion symmetry $\mathcal{I}$ and TRS ($\mathcal{IT}=\mathcal{K}$). 
Therefore, we obtain $\t{Im}[u_{k_F}v_{k_F} v_{12,k_F}]=0$ in inversion symmetric systems and 
no reciprocity appears. This clearly shows that the nonreciprocity in the current mechanism essentially relies on 
the complex nature of wave functions in noncentrosymmetric crystals.

\textbf{Nonreciprocal current in Rice-Mele model.}
We study nonreciprocal current in a representative model of ferroelectrics, Rice-Mele model, by taking into account onsite interaction. We show that $E$-linear band structure change is associated with effective modulation of parameters in the Hamiltonian that is induced by the applied electric field $E$. 

\begin{figure}[tb]
\begin{center}
\includegraphics[width=0.95\linewidth]{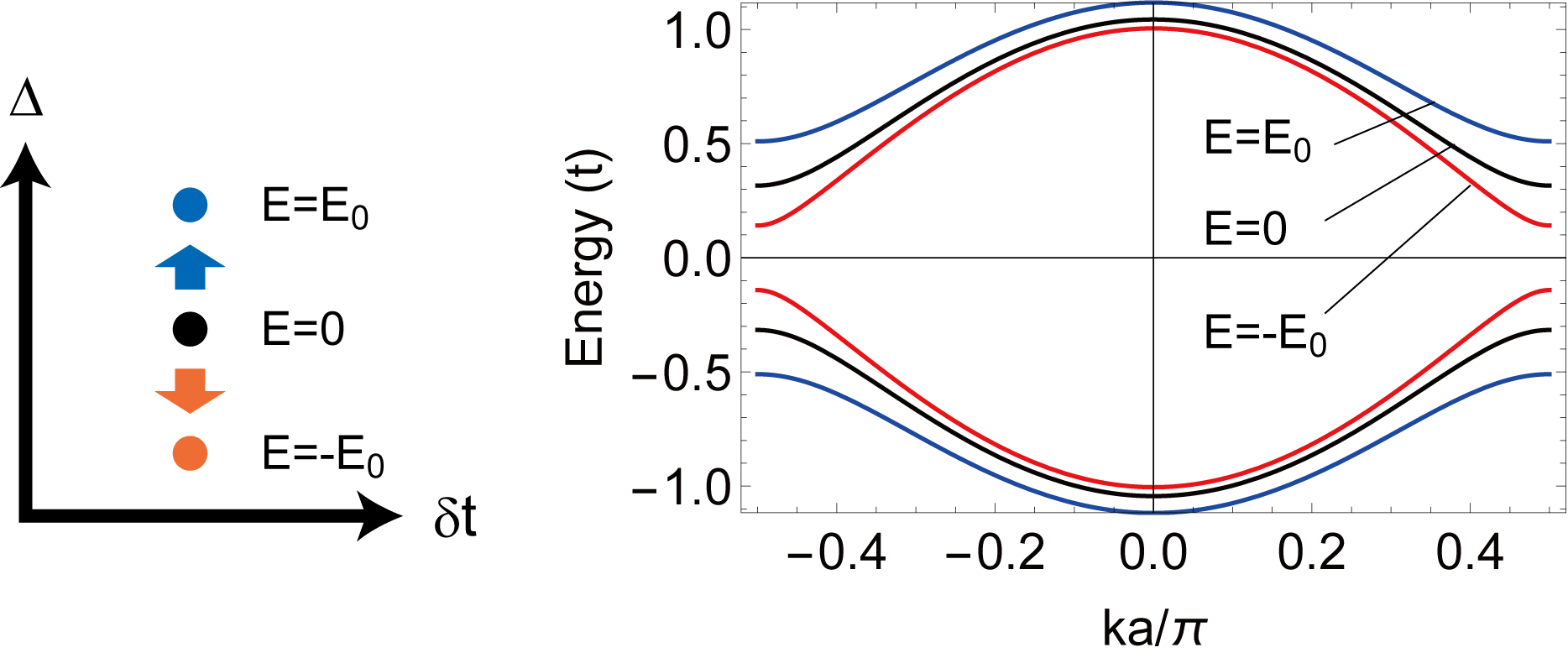}
\caption{\label{fig:RM-mass}
Schematics of the effective parameter change induced by the electric field combined with the electron correlation and the 
associated effective band structures. We adopted parameters $\delta t=0.1, \Delta=0.3$. 
The changes of $\Delta$ from the applied electric fields are $\pm 0.2$.
}
\end{center}
\end{figure}

Rice-Mele model is a representative 1D two-band model 
with broken inversion symmetry,
and is described by a Hamiltonian \cite{Rice-Mele},
\begin{align}
H&=\frac{1}{2} \sum_i (c_{i+1}^\dagger c_i +h.c.) - \frac{\delta t}{2} 
\sum_i (-1)^i (c_{i+1}^\dagger c_i +h.c.) 
\n
&\qquad \qquad 
+ \Delta \sum_i (-1)^{i} c_i^\dagger c_i.
\end{align}
Rice-Mele model is a minimal model for molecular conductors \cite{Su,Nagaosa-Takimoto,Onoda,Nagaosa-Morimoto} and ferroelectric perovskites \cite{Egami}.
In the momentum representation, the Hamiltonian reads
\begin{align}
H&=
\cos \frac{ka}{2}  \sigma_x + \delta t \sin \frac{ka}{2} \sigma_y + \Delta \sigma_z,
\end{align}
where Pauli matrices $\sigma$'s act on two sublattices (A and B) in the unit cell, and $a$ is the lattice constant.
For Rice-Mele model, the wave functions in Eq.~(\ref{eq: Psi k}) are given by $u_k=-\sin \frac \theta 2$ and $v_k=e^{i\phi} \cos \frac \theta 2 $, with the parameters 
$ \theta = \cos^{-1}\frac{\Delta}{|\epsilon_{1,k}|}$ 
and $ \phi = \tan^{-1}\delta t$.
The energy dispersion for the valence band is given by 
\begin{align}
\epsilon_{1,k} &= -\sqrt{\cos^2 \frac{ka}{2} + \delta t^2 \sin^2 \frac{ka}{2} + \Delta^2},
\label{eq: dispersion RM}
\end{align}
and $\epsilon_{2,k} = - \epsilon_{1,k}$ for the conduction band, as shown in the right panel of Fig.~\ref{fig:RM-mass} with black line.

We again consider two copies of Rice-Mele model and introduce the onsite interaction given by
\begin{align}
H_\t{int} &= U \sum_i n_{\uparrow,i} n_{\downarrow,i},
\end{align}
where $\uparrow$ and $\downarrow$ label the two identical copies. 
By focusing on the electronic structure of the $\uparrow$ component, we suppress the label for the two copies for simplicity.

Now we study the nonequilibrium steady state under the electric field $E$ applied along the 1D chain, by using the nonequilibrium Green's functions, Eq.~(\ref{eq:Glesser-inter}).
The electric field is described by the Hamiltonian, 
$H_\t{ele}=- e E a \sum_i i n_i$.
The application of the electric field effectively changes the parameters $\delta t$ and $\Delta$, which can be easily obtained within the Hartree approximation
since the expectation values are directly computed from the lesser component of the Green's function.
From the Hartree term, the occupation of site A is modified as
\begin{align}
\delta n_A 
&=
i \frac{E a}{2}\int \frac{dk}{2\pi}[u_k v_k G^<_{E,21}(k)
+ u_k^* v_k^* G^<_{E,12}(k)] \n
&=
\frac{E a }{\pi |v_{11,k_F}| E_{g,k_F}} \t{Im}\left[u_{k_F}v_{k_F} v_{12,k_F}\right],
\label{eq: delta n A}
\end{align}
(For details of the derivation, see Methods.) 
Similarly, the occupation of site B is modified in the opposite way as
$\delta n_B 
=-\delta n_A$ .
Thus the Hartree term effectively changes the staggered potential $\Delta$ as
\begin{align}
\Delta \to \Delta + \frac{E a U }{\pi |v_{11,k_F}| E_{g,k_F}} 
\t{Im}\left[u_{k_F}v_{k_F} v_{12,k_F}\right].
\label{eq: Delta change RM}
\end{align}
Notice that the change of $\Delta$ is opposite in sign depending on the direction of $E$. This situation is 
schematically illustrated in Fig.~\ref{fig:RM-mass}.
Since the parameter changes are asymmetric with respect to the sign of $E$ 
in the $\delta t-\Delta$ space,
the effective band structure $\epsilon_{1,k}(E)$ becomes
different for the electric fields $+E$ and $-E$.
The nonlinear current in the nonequilibrium steady state is obtained from the conventional Boltzmann equation 
approach for this modified band structure in the presence of $E$.
Namely, the linear conductivity is given by
\begin{align}
\sigma(E) &= 2 \tau |v_F(E)|,
\end{align}
with $E$-dependent Fermi velocity 
$v_F(E)=\partial_k \epsilon_{1,k}(E)|_{k=k_F}$ (where $k_F$ is the Fermi momentum).
The $E$-linear change of the effective band structure leads to $E$-linear term in $v_F(E)$, 
which results in the nonlinear current response $\propto E^2$.
Thus the asymmetry in band structure changes leads to the nonreciprocity of the current 
with respect to the direction of $E$.

The nonreciprocity is quantified by the ratio of the change of electric conductivity 
$\gamma = [\sigma(E)-\sigma(0)]/\sigma(0)$ in the presence of the applied electric field.
We note that $\gamma=\delta J/J$ and the crude approximation is given in Eq.~(\ref{eq: nonreciprocity ratio}). This approximation is also obtained from the $E$-linear change of the parameters in Rice-Mele model in Eq.~(\ref{eq: Delta change RM}) and Eq.~(\ref{eq: dispersion RM}).
Explicit evaluation of Eq.~(\ref{eq: nonlinear delta J}) gives the nonreciprocity ratio of
$\gamma=5 \times 10^{-7}$ for typical parameters of Rice-Mele model ($\delta t=\Delta=0.3t, U=t, k_F=0.1 \pi/a$ along with $t=1\t{ eV}$ and $a=1\t{ \AA}$) and the electric field of $E=10^5\t{ V/m}$.
This order of the nonreciprocity is comparable to those in materials showing magnetochiral anisotropy \cite{Rikken01}, as we will discuss further in the discussion section.

\ \\
\par \ \
\noindent
\textbf{Discussions}
\par

Finally, we give an estimate of the nonreciprocal response induced by the present mechanism for realistic materials.
Typical magnitude of the nonreciprocity 
is determined by $\gamma=\delta J/J $ in Eq.~(\ref{eq: nonreciprocity ratio}).
When the band gap and Coulomb energy are both of the order of 1eV,
the ratio $\delta J/J$ reduces to $eEa/W$,
which is the ratio between the electric potential in the unit cell and the bandwidth.
This allows us to estimate typical nonreciprocity as follows.
We consider the current of 1mA that flows in a wire of the area 1mm$^2$, which amounts to a current density of $j=10^3 \t{A/m}^2$.
For usual metals, conductivity is roughly given by $\sigma \simeq 10^6 \t{A/Vm}$, and hence, the electric field present in the wire is
$E = j/\sigma \simeq 10^{-3} \t{V/m}$.
In this case, the electric potential in the unit cell of $a\simeq 1\AA$ is
$eEa \simeq 10^{-13} \t{eV}$.
Since the bandwidth is typically 1eV, this indicates nonreciprocity ratio is
$\delta J/J \simeq 10^{-13}$.
This should be compared to the typical order of nonreciprocity for materials showing magnetochiral anisotropy.
Bi helix \cite{Rikken01} and molecular solids \cite{Pop14} show the nonreciprocity measured in resistivity change $\delta \rho$ as $\delta \rho/\rho =\gamma' I B$ with $\gamma' \simeq 10^{-3} A^{-1}T^{-1}$.
For $I=1$mA and $B=1$T, the typical nonreciprocity is
$\delta J/J \simeq \delta \rho/\rho \simeq 10^{-6}$.
Thus the nonreciprocity induced by electron correlation is very small for good metals.
On the other hand, we can expect comparable nonreciprocity for doped semiconductors whose conductivity ranges from $10^{-1}\sim 10^{5}$A/Vm.
For example, for the doped Si of $\sigma=10^{-1}$A/Vm and the bandwidth $W\simeq 1$ eV in the presence of the current density $j=10^3 \t{A/m}^2$, we obtain the nonreciprocity of $\delta J/J \simeq 10^{-6}$, which becomes comparable with typical materials showing magnetochiral anisotropy.

Another candidate is the molecular conductor TTF-CA which is a strongly correlated insulator. Of course, our theory for weakly correlated metals is not directly applicable. However, it is interesting to estimate the nonreciprocity ratio anyway, since the carriers in TTF-CA (thermally activated or provided by impurity sites) may be treated as electrons having a Fermi surface, and the Hartree approximation sometimes becomes a good approximation at least for the ground states.  The typical order of electric field that can be applied is $E \simeq 10^5$V/m \cite{Tokura88,Mitani87}. Since the lattice constant is $a \simeq 1$nm, the electric voltage in the unit cell becomes $eEa \simeq 10^{-4}$ eV, and the band width is given by $W\simeq 0.2$ eV. Thus the nonreciprocity ratio can be $10^{-3}$ which may be comparable with that in magnetochiral anisotropy in Bi helix \cite{Rikken01}. We again note that this is a number obtained from a naive application of Eq.~(\ref{eq: nonreciprocity ratio}) to TTF-CA beyond the applicability of our theory, but this suggests that it is an interesting future problem to study TTF-CA as a candidate of strongly correlated materials for nonreciprocity, from both theoretical and experimental points of view.

Our analysis is mostly valid for weakly interacting systems because we adopted Hartree approximation to incorporate the correlation effect. Therefore, the study of nonreciprocal responses of strongly interacting cases remains as an interesting future problem. Meanwhile, our symmetry considerations from generalization of Onsager's theorem suggests that nonreciprocal current response can generally appear in the presence of dissipation and interactions, regardless of the strength of the interaction. We may also note that Hartree approximation sometimes gives a good description for some ground state properties, even for strong $U$ cases, such as magnetically ordered ground states. Our approach may give a good approximation for nonlinear properties of those states, since the nonreciprocal current response is a nonequilibrium property near the ground state under a moderate electric field.



\ \\

\par \ \
\noindent
\textbf{Methods}
\par

\textbf{Derivation of generalized Onsager's theorem.}
In this section, we present general symmetry considerations on the nonreciprocal current response 
with respect to the time reversal symmetry by extending Onsager's relationship to nonlinear current.

We consider a system of noninteracting electrons that are described by Green's function in the Lehmann representation,
\begin{align}
G_{ab}(i\omega_n)&=
e^{\beta \Omega} \sum_{\alpha,\beta}
\braket{\alpha|c_a|\beta} \braket{\beta|c_b^\dagger|\alpha} \frac{e^{-\beta E_\alpha}+e^{-\beta E_\beta}}{i\omega_n+E_\alpha-E_\beta},
\label{eq: lehmann rep}
\end{align}
where $\ket{\alpha}$ is a many-body state that satisfies $\hat{H}\ket{\alpha}=E_\alpha \ket{\alpha}$ with the many-body Hamiltonian $\hat{H}$, $\beta$ is the inverse temperature, $e^{-\beta \Omega}=\t{Tr}[e^{-\beta \hat{H}}]$, and $c_a$ and $c_a^\dagger$ are annihilation and creation operators of an electron with a single particle state $a$. (Here $\alpha,\beta$ are labels for many-body states, whereas $a,b$ are labels for single particle states.)
We write the current operator $\hat{v_i}$ along the $i$th direction as
\begin{align}
\hat{v_i}=\sum_{ab} (v_i)_{ab} c^\dagger_a c_b,
\label{eq: v op many body}
\end{align}
where $v$ is a matrix for a velocity operator in the single particle representation.

In the linear response, Onsager's relationship indicates that the conductivity $\sigma_{ij}$ 
is constrained as
\begin{align}
\sigma_{ij} &= \sigma_{ji},
\end{align}
in the presence of time reversal symmetry \cite{Onsager-reciprocal}.
This relationship is derived by considering the time reversal transformation in the Kubo formula 
for the linear conductivity,
\begin{align}
\sigma_{ij}(i\omega_n) &= 
\frac{1}{\omega_n \beta} \sum_{i\omega_m} \t{tr}[
v_i G(i\omega_m + i \omega_n)  
 v_j G(i\omega_m)
],
\end{align}
where $i\omega_n, i\omega_m$ are Matsubara frequencies, and $\t{tr}$ is a trace over single particle states (labeled by $a,b$).
The time reversal symmetry, $\mathcal{T}=K$, indicates
\begin{align}
G(i\omega_m) &= G^T(i\omega_m), \\
v_i &= -v_i^T.
\end{align}
These actions of $\mathcal{T}$ in the many-body representation are obtained by using $\mathcal{T}\ket{\alpha}=(\ket{\alpha})^*$ in Eq.~(\ref{eq: lehmann rep}) and Eq.~(\ref{eq: v op many body}). (We note that this is closely related to symmetry constraint in a single particle Hamiltonian,
$H(k) = H^T(-k)$, in the momentum representation.)
By using these relationships, the Kubo formula can be rewritten as
\begin{align}
\sigma_{ij}(i\omega_n) 
&= 
\frac{1}{\omega_n \beta}\sum_{i\omega_m} \t{tr}[
v_i^T G^T(i\omega_m + i \omega_n) v_j^T G^T(i\omega_m)
] 
\n
&=
\frac{1}{\omega_n \beta}\sum_{i\omega_m} \t{tr}[
v_j G(i\omega_m + i \omega_n) v_i G(i\omega_m)
]
\n
&=\sigma_{ji}(i\omega_n),
\end{align}
and leads to the Onsager's relationship. Here we rewrote the trace in the reverse order 
in the second line and used the fact that the transposition in the trace does not change its value.

Next we study how Onsager's theorem can be extended to nonlinear 
current responses. We consider the second order current response,
\begin{align}
J_i(\omega_1 + \omega_2) &= \sigma_{ijj}(\omega_1, \omega_2) E_j(\omega_1)E_j(\omega_2).
\end{align}
The nonlinear conductivity $\sigma_{ijj}(i\omega_{n_1},i\omega_{n_2})$ has a 
contribution from a triangle diagram which is given by
\begin{widetext}
\begin{align}
\sigma^\t{tr}_{ijj}(i\omega_{n_1},i\omega_{n_2})
&=
\frac{1}{\omega_{n_1}\omega_{n_2} \beta}
\sum_{i\omega_m} \t{tr}[
v_j G(i\omega_m + i \omega_{n_1}) v_j G(i\omega_m + i \omega_{n_1} 
+ i \omega_{n_2}) v_i G(i\omega_m)
],
\end{align}
since there are no vertex corrections for noninteracting systems.
The time reversal symmetry indicates that 
\begin{align}
\sigma^\t{tr}_{ijj}(i\omega_{n_1},i\omega_{n_2})
&=
\frac{-1}{\omega_{n_1}\omega_{n_2} \beta}
\sum_{i\omega_m}\t{tr}[
v_j^T G^T(i\omega_m + i \omega_{n_1}) v_j^T G^T(i\omega_m + i \omega_{n_1} 
+ i \omega_{n_2}) v_i^T G^T(i\omega_m)
] \n
&=
\frac{-1}{\omega_{n_1}\omega_{n_2} \beta}
\sum_{i\omega_m} \t{tr}[
v_j G(i\omega_m - i \omega_{n_2}) v_j G(i\omega_m - i \omega_{n_1} 
- i \omega_{n_2}) v_i G(i\omega_m )
] 
\n
&=
-\sigma^\t{tr}_{ijj}(-i\omega_{n_2},-i\omega_{n_1})
.
\end{align}
\end{widetext}
Naively, this seems to suggest that the nonlinear conductivity $\sigma_{ijj}(\omega_1,\omega_2)$ 
vanishes in the dc limit ($\omega_1\to 0$ and $\omega_2\to 0$).
However, we notice that nonlinear conductivity with Matsubara frequencies in the upper half plane is 
transformed to that with Matsubara frequencies in the lower half plane.
Since the real axis is a branch cut in the complex $\omega$ plane, the analytic continuation of 
$i\omega_n \to 0$ for the two quantities, $\sigma_{ijj}(i\omega_1,i\omega_2)$ and 
$\sigma_{ijj}(-i\omega_2,-i\omega_1)$, lead to different results in general.
This indicates that relationship similar to the Onsager's relation does not necessarily constrain the 
dc nonlinear conductivity to vanish.
Instead, this extended Onsager's relation indicates that nonreciprocal current necessarily involves irreversibility such as dissipation and relaxation.

In a similar manner, we can also derive an extended Onsager's relation for shift current. Shift current is dc current induced by optical absorption above the band gap and photoexcitation of electron-hole pairs that have finite polarization \cite{Sipe,Morimoto-shift16}.
It is described by a nonlinear current response,
$J_i(\omega_1+\omega_2)= \sigma_{ijj}^\t{shift}(\omega_1,\omega_2) E(\omega_1) E(\omega_2)$ with $\omega_2 \approx -\omega_1$.
The nonlinear conductivity $\sigma^\t{shift}$ has two contributions as $\sigma^\t{shift}=\sigma^\t{tr}(\omega_1,\omega_2) + \sigma^\t{bubble}(\omega_1,\omega_2)$, where the latter piece is a correlation function of paramagnetic current $\hat{v}_i$ and diamagnetic current $\hat{v}_\t{dia,ij}\equiv \sum_{ab} (v_{\t{dia},ij})_{ab} c_a^\dagger c_b$ \cite{Kim17}. Time reversal symmetry leads to the same relation,
\begin{align}
\sigma_{ijj}^\t{shift}(\omega_1,\omega_2) = - \sigma_{ijj}^\t{shift}(-\omega_2,-\omega_1),
\end{align}
since $\sigma^\t{bubble}$ also obeys the same transformation law under the TRS with $\sigma^\t{tr}$ as follows.
In the momentum representation, matrix elements for diamagnetic current are given by
\begin{align}
v_{\t{dia},ij}=\frac{\partial v_j}{\partial k_i}.
\end{align}
Accordingly, TRS constrains diamagnetic current operator as
\begin{align}
v_\t{dia}=v_\t{dia}^T,
\end{align}
due to an extra $k$ derivative.
The nonlinear conductivity for shift current is written as
\begin{align}
&\sigma_{ijj}^\t{bubble}(i\omega_{n_1},i\omega_{n_2}) \n
&= 
\frac{1}{\omega_{n_1}\omega_{n_2} \beta} \sum_{i=1,2} \sum_{i\omega_m} \t{tr}[
v_j G(i\omega_m + i \omega_{n_i})  
v_{\t{dia},ij} G(i\omega_m)
].
\end{align}
Under TRS, this transforms as
\begin{align}
&\sigma_{ijj}^\t{bubble}(i\omega_{n_1},i\omega_{n_2}) \n&= 
-\frac{1}{\omega_{n_1}\omega_{n_2} \beta} \sum_{i=1,2}\sum_{i\omega_m} \t{tr}[
v_i^T G^T(i\omega_m + i \omega_{n_i}) v_{\t{dia},ij}^T G^T(i\omega_m)
] 
\n
&=
-\frac{1}{\omega_{n_1}\omega_{n_2} \beta} \sum_{i=1,2}\sum_{i\omega_m} \t{tr}[
v_i G(i\omega_m - i \omega_{n_i}) v_{\t{dia},ij} G(i\omega_m)
]
\n
&=-\sigma_{ijj}^\t{bubble}(-i\omega_{n_2}, - i\omega_{n_1}),
\end{align}
where we used the symmetry between $i\omega_{n_1}$ and $i\omega_{n_2}$ to fit the transformation law with that for $\sigma^\t{tr}$.
Therefore, nonzero shift current also requires irreversibility that introduces a branch cut at the real axis in the $\omega$ space and makes two limits $\omega \to \pm i 0$ different. In this case, the irreversibility comes from optical transition and creation of electron hole pairs across the band gap.

\textbf{Keldysh Green's function.}
In this section, we summarize basic notations of Keldysh Green's functions that we need for our discussion~\cite{Jauho94,Kamenev,Aoki-RMP14}.
In the Keldysh Green's function formalism, we consider the Keldysh component of the Green's function in addition to the retarded and advanced Green's function.
Keldysh component describes the electron occupation in the nonequilibrium state, while the retarded and advanced components describe the spectrum of the system.
The Dyson equation for the Green's function is given by
\begin{align}
\begin{pmatrix}
G^R & G^K \\
0 & G^A
\end{pmatrix}^{-1}
&=
\omega-H -
\begin{pmatrix}
\Sigma^R & \Sigma^K \\
0 & \Sigma^A
\end{pmatrix},
\end{align}
with the Hamiltonian $H$.

In the thermal equilibrium, the Keldysh Green's function is obtained by solving the Dyson equation. We suppose that the system is weakly coupled to a heat bath with broad spectrum, which determines the electron distribution of the system. The coupling to the heat bath (such as electron reservoirs) is described by the self energy given by \cite{Jauho94}
\begin{align}
\begin{pmatrix}
\Sigma^R & \Sigma^K \\
0 & \Sigma^A
\end{pmatrix}
&=
i \Gamma
\begin{pmatrix}
- \frac 1 2 & 2f-1 \\
0 & \frac 1 2
\end{pmatrix},
\label{eq: Sigma heat bath}
\end{align}
where $\Gamma$ is the strength of the coupling to the bath, and $f(x)=1/[1+\exp(x/k_B T)]$ is the Fermi distribution function with the temperature $T$.

The observables in the nonequilibrium steady state is obtained from Keldysh Green's function.
We define the lesser component of the Green's function as
\begin{align}
G^<(\omega, k) \equiv \frac{1}{2}(G^K-G^R+G^A).
\end{align}
By using $G^<$, we can write the expectation value of a general fermion bilinear as
\begin{align}
\braket{c_j^\dagger c_i} = -i \int \frac{d\omega}{2\pi} G_{ij}^< (\omega).
\label{eq: expectation value}
\end{align}
The lesser Green's function is concisely obtained from the equation
\begin{align}
G^<=G^R \Sigma^< G^A,
\end{align}
where the lesser component of the self energy encodes the information of the electron distribution and is given by
\begin{align}
\Sigma^<(\omega,k) \equiv \frac{1}{2}(\Sigma^K-\Sigma^R+\Sigma^A) = i\Gamma f(\omega).
\end{align}

\textbf{Keldysh Green's function under the applied electric field.}
In this section, we study the nonequilibrium electron distribution realized under the applied electric field.
We compute the $E$-linear part of the lesser Green's function $G_E^<$ in Eq.~(\ref{eq: G G0 GE}) in the gauge invariant formulation.
In doing so, we use the diagram in Fig.~\ref{fig:diagrams}(b) to specify the form of self energy $\Sigma^<_E$ in Eq.~(\ref{eq: G G0 GE}). 
(We note that the electron interaction in Fig.~\ref{fig:diagrams}(a) does not change electron distribution and does not contribute to $\Sigma^<_E$. Furthermore, it turns out in the end that the contribution of impurity scattering to $\Sigma^<_E$ is actually negligible under TRS.)
Specifically, we consider the delta function type impurity [$V(r)=u\delta(r-r_0)$ with density $n$]. 
In the second-order Born approximation, the self energy is given  by 
\begin{align}
\Sigma_E(\omega, k)=n u^2 \int \frac{dk}{2\pi} G_E(\omega, k),
\label{eq: sigma impurity}
\end{align}
which corresponds to the diagram in Fig.~\ref{fig:diagrams}(b).
In the right hand side, $G_E$ denotes the bare Green's function that does not include the effect of impurity scattering.
We note that the impurity scattering also modifies the self energy $\Sigma_0$ in the zeroth order in $E$, but this correction only changes the coupling $\Gamma$ in Eq.~(\ref{eq: Sigma heat bath}) and can be absorbed by redefining $\Gamma$ accordingly.

The current response in the nonequilibrium steady state under the electric field is captured by the lesser Green's function $G_E^<$ which gives a contribution linear in $E$.
We consider a multiband system and suppose that the Bloch wave functions are given by $\Psi_{i,k}$ which satisfy $H \Psi_{i,k} = \epsilon_{i,k} \Psi_{i,k}$ with the energy dispersion $\epsilon_{i,k}$ (where $i$ is the band index).
First we start with the intraband component of $G_{E,ii}^<$ for the band $i$ (where we omit the band index $i$ in the following, for simplicity).
By assuming that $G_0$ and $v$ have a single component, equation~(\ref{eq:GE}) gives
\begin{align}
G_E^< 
&=
G_0^R \left[ \Sigma_E^< 
+ \frac{i}{2} 
 \left( 
(\partial_{\omega} \Sigma^<) G_0^A  v_k
-
v_k G_0^R (\partial_{\omega} \Sigma^<)
\right) \right] G_0^A 
\n
&=
\frac{\Sigma_E^<}{(\omega-\epsilon_k)^2+\frac{\Gamma^2}{4}}
+ 
\frac{i\Gamma^2 v_k \delta(\omega-\epsilon_F)}{2[(\omega-\epsilon_k)^2
+\frac{\Gamma^2}{4}]^2}
,
\label{eq: GE<0}
\end{align}
where we used $\partial_\omega f(\omega)= - \delta(\omega-\epsilon_F)$.
This expression is simplified by using the relationship
\begin{align}
\frac{1}{[(\epsilon_F-\epsilon_k)^2+ \frac{\Gamma^2}{4}]^n}
&=\frac{2\pi (2n-2)!}{[(n-1)!]^2 \Gamma^{2n-1}}\sum_{k_{F,i}}\frac{1}{|v_k|} 
\delta(k-k_{F,i}),
\label{eq: delta relationships}
\end{align}
that holds for a positive integer $n$, and the Fermi momenta $k_{F,i}$,
where we only keep the leading order in terms of $1/\Gamma$. 
By using Eq.~(\ref{eq: GE<0}) with Eq.~(\ref{eq: sigma impurity}), the impurity scattering gives rise to $\Sigma_E^<(\omega)$ 
given by
\begin{align}
\Sigma_E^<(\omega)
&=
 i nu^2 \frac{2\pi}{\Gamma} \delta(\omega-\epsilon_F) \n
&\quad \times 
\frac{
\sum_{k_{F,i}}\frac{v_k}{|v_k|} \delta(k-k_{F,i})
}{
1- nu^2 \frac{2\pi}{\Gamma} \sum_{k_{F,i}}\frac{1}{|v_k|} \delta(k-k_{F,i})
}
.
\end{align}
Here, the numerator in the right hand side vanishes since the TRS leads to 
$ \sum_{k_{F,i}}\frac{v_k}{|v_k|} =0 $, and hence, $\Sigma_E^<(\omega)=0$ follows. 
Thus we obtain
\begin{align}
G_E^< &= \frac{2\pi i}{\Gamma} \delta(\omega-\epsilon_F) \sum_{k_{F,i}}\frac{v_k}{|v_k|} 
\delta(k-k_{F,i}).
\end{align}
This change of the lesser Green's function linear in $E$ describes the effect of the applied electric field 
where the electron occupation is shifted in the momentum space as $k \to k + \tau E$ near the Fermi 
surface (with $\tau = 2\pi/\Gamma$).
This corresponds to the picture from the semiclassical Boltzmann equation as illustrated in 
Fig.~\ref{fig: schematics}(a).

Next we consider the interband component, $G_{E,12}^<$, by focusing on the valence and conduction bands which are labeled by 1 and 2, respectively.
Equation~(\ref{eq:GE}) gives
\begin{align}
G_{E,12}^< &=
G_{0,11}^R \Big[ \Sigma_{E,12}^< 
+ \frac{i}{2} 
 \big( 
(\partial_{\omega} \Sigma^<_{11}) G_{0,11}^A  v_{k,12}
\n
&\hspace{8em}
-
v_{k,12} G_{0,22}^R (\partial_{\omega} \Sigma^<_{22})
\big) \Big] G_{0,22}^A .
\end{align}
We assume that the Fermi energy is located within the band 1 and does not cross the band 2.
In this case, the second term in the right hand side reduces to
\begin{align}
&G_{E,12}^< - G_{0,11}^R \Sigma_{E,12}^< G_{0,22}^A 
\n
&=
\frac{\Gamma v_{12,k}}{2} \delta(\omega-\epsilon_F)
[G_{0,11}^R G_{0,11}^A G_{0,22}^A - G_{0,11}^R G_{0,22}^R G_{0,22}^A]
\n
&=
- \sum_{k_{F,i}}
\frac{\pi v_{12,k}}{|v_{11,k}| E_{g,k}} \delta(\omega-\epsilon_F) \delta(k-k_{F,i}) 
,
\end{align}
with $E_{g,k}=\epsilon_{2,k}-\epsilon_{1,k}$,
where we only kept the leading term with respect to $1/E_{g,k}$.
(Here we used Eq.~(\ref{eq: delta relationships}) for $G_{0,11}^R G_{0,11}^A$ and discarded the second term.)
Since the right hand side is inversely proportional to the band gap $E_{g,k}$, the 
self energy $\Sigma_{E,12}^<$ obtained from Eq.~(\ref{eq: sigma impurity}) is proportional 
to $\Gamma/E_{g,k_F}$, which is negligible in the left hand side of the above equation given that $G_{0,22}^A \propto 1/E_{g,k_F}$.
Therefore the lesser part of the Green's function is given by
\begin{align}
G_{E,12}^< &=
- \sum_{k_{F,i}}
\frac{\pi v_{12,k}}{|v_{11,k}| E_{g,k}} \delta(\omega-\epsilon_F) \delta(k-k_{F,i})
.
\end{align}
We note that $G_{E,21}^<$ is obtained from the relationship
\begin{align}
G_{E,21}^< &= - (G_{E,12}^<)^*,
\end{align}
as a consequence of the hermiticity of expectation values in Eq.~(\ref{eq: expectation value}).


\textbf{Effective band dispersion with impurity scattering.}
In this section, we study the effective band dispersion in the presence of $E$ and impurity scattering  by looking at $\Sigma_E^R$.
We show that impurity scattering is insufficient for nonreciprocal current response because the change of the band dispersion turns out to be the same for positive and negative electric fields.

From Eq.~(\ref{eq:GE}), the retarded part of the equation for $G_E$ reads
\begin{align}
G_E^R=G_0^R \left[
\Sigma_E^R + \frac{i}{2} 
\left( 
G_0^R (-v_k)
-
(-v_k) G_0^R 
\right) \right] G_0^R,
\label{eq: app G E R}
\end{align}
with $v_k=\partial_k H$, where we used $\partial_\omega \Sigma^R=0$.
For simplicity, we consider a two-band system, where the Green's function is given by 
\begin{align}
G_{0,ij}^R=
\frac{1}{\omega-\epsilon_i+i \frac \Gamma 2} \delta_{ij},
\end{align}
where $i,j=1,2$ are labels for valence and conduction bands, respectively.
For the diagonal components, we obtain
\begin{align}
G_{E,ii}^R=G_{0,ii}^R \Sigma_{E,ii}^R G_{0,ii}^R,
\end{align}
since the second term in Eq.~(\ref{eq: app G E R}) vanishes trivially.
The diagonal part of the self energy is momentum independent and vanishes as
\begin{align}
\Sigma_{E,ii}^R (\omega) &= n u^2 \int \frac{dk}{2\pi} G_{E,ii}^R \n
&= 
\left[n u^2  \int \frac{dk}{2\pi} \frac{1}{(\omega-\epsilon_i(k)+i \frac \Gamma 2)^2} \right]
\Sigma_{E,ii}^R(\omega) =0.
\end{align}
Off-diagonal part is determined from
\begin{align}
&G_{E,21}^R - G_{0,22}^R \Sigma_{E,21}^R G_{0,11}^R
\n
&= 
- \frac{i}{2} G_{0,22}^R 
\left( 
G_{0,22}^R v_{k,21}
-
v_{k,21} G_{0,11}^R 
\right) G_{0,11}^R
\n
&=
- \frac{i v_{k,21} (\epsilon_1-\epsilon_2)}{2 (\omega-\epsilon_1+i \frac \Gamma 2)^2(\omega-\epsilon_2+i \frac \Gamma 2)^2}.
\end{align}
By integrating over the momentum, we obtain
\begin{align}
&\left(1-nu^2 \int \frac{dk}{2\pi} G_{0,22}^R G_{0,11}^R \right) \Sigma_{E,21}^R(\omega) 
\n
&=
- n u^2 \int \frac{dk}{2\pi}
 \frac{i v_{k,21} (\epsilon_1-\epsilon_2)}{2 (\omega-\epsilon_1+i \frac \Gamma 2)^2(\omega-\epsilon_2+i \frac \Gamma 2)^2},
\end{align}
which leads to nonzero $\Sigma_{E,21}^R$ in general.
Therefore, the effective Hamiltonian is given by
\begin{align}
H=
\begin{pmatrix}
\epsilon_1 & \frac{E}{2} \Sigma_{E,12}^R \\
\frac{E}{2} \Sigma_{E,21}^R & \epsilon_2
\end{pmatrix},
\label{eq: modified H impurity}
\end{align} 
and the effective band structure of the valence band in the presence of $E$ is obtained by diagonalizing $H$ as
\begin{align}
\widetilde \epsilon_1=\frac{\epsilon_1+\epsilon_2}{2}-\sqrt{\frac{(\epsilon_2-\epsilon_1)^2}{4}+\frac{E^2 |\Sigma_{E,21}^R|^2}{4}}.
\end{align}
This is an even function with respect to $E$; the effective band structure depends on the strength of electric field $|E|$, but is independent of the direction of the applied field. 
Therefore, no reciprocal current appears when we use Boltzmann equation approach based on this modified band structure.

We note that this conclusion is not changed even when we treat the impurity scattering by self-consistent Born approximation. In the self-consistent Born approximation, $G_E$ in Eq.~(\ref{eq: sigma impurity}) is taken as a full Green's function including the effect impurity scattering. In this case, the self energy $\Sigma_E$ is obtained by repeating the above calculation and taking convergence. In the every step of the repetition, the energy dispersion is modified as Eq.~(\ref{eq: modified H impurity}) and still gives a symmetric dispersion in $k$. After repeating this many times, the dispersion remains symmetric in $k$.
Therefore, self-consistent treatment of impurity scattering still gives no nonreciprocal current response.

\textbf{Electron-electron interaction in two-band model.}
In this section, we derive Eq.~(\ref{eq: Sigma E 11 R}) for the self energy that arises from the electron-electron interaction in the case of a two-band model. We also derive Eq.~(\ref{eq: delta n A}) for the expectation value of density operators. These expressions are obtained by using the momentum space representation of the interaction Hamiltonian.

We consider the onsite interaction that is given by
\begin{align}
H_\t{int}&=U \sum_n 
(n_{A,\uparrow,n} n_{A,\downarrow,n} + n_{B,\uparrow,n} n_{B,\downarrow,n}),
\end{align}
with the site index $n$.
Expressing the Hartree contribution to the self energy requires momentum representations of the density operators $n_{A,i}$ and $n_{B,i}$, where we omit the indices for two copies ($\uparrow$ and $\downarrow$) since the expressions are identical for two copies.
For the wave functions in Eq.~(\ref{eq: Psi k}), 
the creation operators of Bloch states are written as
\begin{align}
c_{1,k}^\dagger &= \frac{1}{\sqrt N} \sum_n e^{ikn}(u_k c_{A,n}^\dagger + v_k c_{B,n}^\dagger), \\
c_{2,k}^\dagger &= \frac{1}{\sqrt N} \sum_n e^{ikn}(-v_k^* c_{A,n}^\dagger + u_k^* c_{B,n}^\dagger), 
\end{align}
where $N$ is the system size.
By using inverse Fourier transformation, the creation operators in the site basis are expressed with Bloch states as
\begin{align}
c_{A,n}^\dagger &= \frac{1}{\sqrt N} \sum_k e^{-ikn}(u_k^* c_{1,k}^\dagger - v_k c_{2,k}^\dagger), \\
c_{B,n}^\dagger &= \frac{1}{\sqrt N} \sum_k e^{-ikn}(v_k^* c_{1,k}^\dagger + u_k c_{2,k}^\dagger), 
\end{align}
where $k$ runs momenta in the first Brillouin zone (e.g., $k=2\pi j/Na$ for $j=0,\ldots, N-1$ with lattice constant $a$).
Now the density operators are given by
\begin{widetext}
\begin{subequations}
\begin{align}
c_{A,n}^\dagger c_{A,n} &=
\frac{1}{N} \sum_{k_1,k_2} e^{-i(k_1-k_2)n} \Big[
u_{k_1} u_{k_2}^* c_{1,k_1}^\dagger c_{1,k_2}  
+ v_{k_1} v_{k_2}^* c_{2,k_1}^\dagger c_{2,k_2}  
- u_{k_1} v_{k_2} c_{1,k_1}^\dagger c_{2,k_2} 
- v_{k_1}^* u_{k_2}^* c_{2,k_1}^\dagger c_{1,k_2} 
\Big], \\
c_{B,n}^\dagger c_{B,n} &=
\frac{1}{N} \sum_{k_1,k_2} e^{-i(k_1-k_2)n} \Big[
v_{k_1} v_{k_2}^* c_{1,k_1}^\dagger c_{1,k_2}  
+ u_{k_1}^* u_{k_2} c_{2,k_1}^\dagger c_{2,k_2}  
+ v_{k_1} u_{k_2} c_{1,k_1}^\dagger c_{2,k_2} 
+ u_{k_1}^* v_{k_2}^* c_{2,k_1}^\dagger c_{1,k_2} 
\Big].
\end{align}
\label{eq:density ops}
\end{subequations}
The retarded part of the self energy is given by \cite{Hanai16}
\begin{align}
\Sigma_{E,m_1 m_2}^R (\omega,k) &= - \frac{i}{N} \sum_{k'} \int \frac{d\omega'}{2\pi} \left[ 
U_{(m_1,k)(m_3,k');(m_2,k)(m_4,k')}-U_{(m_3,k')(m_1,k);(m_2,k)(m_4,k')}
\right] G_{E,m_3 m_4}^< (\omega',k'),
\label{eq: Sigma E R}
\end{align}
by using the momentum space representation for the interaction
$H_\t{int}$ which is given by
\begin{align}
H_\t{int}&= - \frac{1}{2 N} \sum_{m_1,m_2,m_3,m_4}\sum_{k_1,k_2,k_3,k_4} \delta(k_1+k_2-k_3-k_4) 
U_{(m_1, k_1)(m_2, k_2); (m_3, k_3) (m_4, k_4)} c_{m_1, k_1}^\dagger c_{m_2,k_2}^\dagger c_{m_3,k_3} c_{m_4,k_4},
\end{align}
with the band index $m_i$.
The first term in Eq.~(\ref{eq: Sigma E R}) is the Hartree term, and the second is the Fock term.
The momentum representations of the interaction that are relevant for the Hartree term contributing to $\Sigma_{E,11}^R$ are given by
\begin{align}
U_{(1,k),(1,k');(1,k),(1,k')} &=
U[|u_k|^2 |u_{k'}|^2 + |v_k|^2 |v_{k'}|^2], \\
U_{(1,k),(2,k');(1,k),(2,k')} &=
U[|u_k|^2 |v_{k'}|^2 + |v_k|^2 |u_{k'}|^2], \\
U_{(1,k),(1,k');(1,k),(2,k')} &=
U(-|u_k|^2 + |v_k|^2) u_{k'}v_{k'}, \\
U_{(1,k),(2,k');(1,k),(1,k')} &=
U(-|u_k|^2 + |v_k|^2) u^*_{k'}v^*_{k'}. 
\end{align}
By using Eq.~(\ref{eq: Sigma E R}), 
the self energy $\Sigma_{E,11}^R$ is written as
\begin{align}
\Sigma_{E,11}^R(k)
=
-i \frac{U}{N} \sum_{k'} \int \frac{d\omega}{2\pi} 
&\Big\{(|u_k|^2 |u_{k'}|^2 + |v_k|^2 |v_{k'}|^2) G_{E,11}^<(k') 
+
(|u_k|^2 |v_{k'}|^2 + |v_k|^2 |u_{k'}|^2) G_{E,22}^<(k') 
\n
& 
+ (-|u_k|^2 + |v_k|^2) 
[ u_{k'}v_{k'} G_{E,12}^<(k') + u_{k'}^* v_{k'}^* G_{E,21}^<(k') ]
\Big\}.
\end{align}
The first two terms in the integral vanishes due to TRS. Specifically, $G_{E,ii}(k)$ is an odd function of $k$ due to TRS as in Eq.~(\ref{eq:GElesser}), and $|u_k|^2$ and $|v_k|^2$ are even functions of $k$, which indicates that the first two terms vanish after integrating over $k'$.
Thus we end up with
\begin{align}
\Sigma_{E,11}^R(k)
= ia U \int \frac{d\omega}{2\pi} \frac{dk}{2\pi} (|u_k|^2 - |v_k|^2) 
[ u_{k'}v_{k'} G_{E,12}^<(k') + u_{k'}^* v_{k'}^* G_{E,21}^<(k') ],
\end{align}
where we replaced the sum $\sum_{k'}$ with the integral $N a\int \frac{dk}{2\pi}$.

Next, we derive the changes of the density $\delta n_A$ and $\delta n_B$ caused by the electric field $E$.
By using Eq.~(\ref{eq:density ops}) and Eq.~(\ref{eq: expectation value}), the change of the density at A site is given by
\begin{align}
\delta n_A &= 
-i \frac{E a}{2} \int \frac{dk}{2\pi} [|u_k|^2 G^<_{E,11}(k) + |v_k|^2 G^<_{E,22}(k) - u_k v_k G^<_{E,21}(k)
- u_k^* v_k^* G^<_{E,12}(k)].
\end{align}
Since the first and second terms vanish due to TRS, we obtain
\begin{align}
\delta n_A &=
i \frac{E a}{2}\int \frac{dk}{2\pi}[u_k v_k G^<_{E,21}(k)
+ u_k^* v_k^* G^<_{E,12}(k)].
\end{align}
Similarly, the change of the density at A site is given by
\begin{align}
\delta n_B &=
-i \frac{E a}{2}\int \frac{dk}{2\pi}[u_k v_k G^<_{E,21}(k)
+ u_k^* v_k^* G^<_{E,12}(k)],
\end{align}
which is opposite in sign compared to $\delta n_A$.

\end{widetext}

\textbf{Acknowledgements.}
This work was supported by
the Gordon and Betty Moore Foundation's EPiQS Initiative Theory Center Grant (TM),
and by Grants-in-Aid for Scientific Research from the Ministry of
Education, Science, Sports and Culture No. 24224009 and 25400317, CREST, Japan Science and Technology (grant no. JPMJCR16F1), and ImPACT Program of Council for Science, Technology and Innovation (Cabinet office, Government of Japan, 888176) (NN).

\bibliographystyle{naturemag.bst}
\bibliography{NonReciprocal.bib}

\end{document}